\documentclass[aps,prd,twocolumn,flushleft,amsmath,amssymb,nofootinbib]{revtex4}
\usepackage{txfonts}
\usepackage{subfig}
\usepackage{graphicx}
\usepackage{dcolumn} 
\usepackage{bm}      
\usepackage[active]{srcltx} 
\usepackage{hyperref} 
\usepackage{amsmath}
\usepackage{amssymb}
\usepackage{stmaryrd}
\usepackage{amsfonts}
\usepackage{mathrsfs}

\newcommand{\be}{\begin{equation}}
\newcommand{\ee}{\end{equation}}
\newcommand{\ba}{\begin{eqnarray}}
\newcommand{\ea}{\end{eqnarray}}

\begin{document}

\title{Action principle for the connection  dynamics of scalar-tensor theories}
\author{Zhenhua Zhou\footnote{dtplanck@163.com}, Haibiao Guo, Yu Han and Yongge Ma\footnote{mayg@bnu.edu.cn}}
\affiliation{Department of Physics, Beijing Normal University, Beijing 100875, China}

\begin{abstract}
A first-order action for scalar-tensor theories of gravity is proposed. The Hamiltonian analysis of
the action gives the desired connection dynamical formalism, which was derived
from the geometrical dynamics by canonical transformations. It is shown that
this connection formalism in Jordan frame is equivalent to the alternative connection formalism in Einstein frame.
Therefore, the action principle underlying loop
quantum scalar-tensor theories is recovered.\\

\end{abstract}

\maketitle

\section{Introduction}\label{sec:introduction}

Modified gravity theories have recently received increased attention in issues related to
the "dark Universe" and nontrivial tests on gravity beyond general relativity (GR). Since 1998,
a series of independent astronomic observations implied that our Universe is currently undergoing
a period of accelerated expansion \cite{Friemann}. This causes the "dark energy" problem in the framework of GR.
It is thus reasonable to consider the possibility that GR is not a valid theory of gravity on a
galactic or cosmological scale. A simple and typical modification of GR is the so-called $f(R)$
theory of gravity \cite{So}. Besides $f(R)$ theories, a well-known competing relativistic theory of gravity
was proposed by Brans and Dicke in 1961 \cite{Brans}, which is apparently compatible with Mach's principle.
To represent a varying "gravitational constant", a scalar field is nonminimally coupled to the metric
in Brans-Dicke theory. To be compared with the observational results within the framework of broad class
of theories, the Brans-Dicke theory was generalized by Bergmann \cite{Berg} and Wagoner \cite{Wag} to general scalar-tensor
theories (STT). Scalar-tensor modifications of GR are also popular in unification schemes such as
string theory (see, e.g., \cite{str1} \cite{str2} \cite{str3}). Note that the metric $f(R)$ theories and Palatini $f(R)$ theories are
equivalent to the special kinds of STT with the coupling parameter $\omega=0$ and $\omega=-\frac{3}{2}$
respectively \cite{So}, while the original Brans-Dicke theory is the particular case of constant $\omega$ and
vanishing potential of $\phi$.

In the past two decades, a nonperturbative quantization of GR, called loop quantum gravity (LQG),
has matured \cite{lqg1} \cite{lqg2} \cite{lqg3} \cite{lqg4}.
It is remarkable that both $f(R)$ theories and STT can be nonperturbatively quantized by extending
the LQG techniques \cite{ZXD1} \cite{ZXD2} \cite{ZXD3}. Thus LQG is extended to more general metric theories
of gravity \cite{Ma,ZhM}. The background independent quantization method relies on the key observations that these
theories can be  cast into the connection dynamical formulations with the structure group $SU(2)$.
The connection dynamical formulation of $f(R)$ theories and STT were obtained by canonical transformations
from their geometrical dynamics \cite{ZXD1} \cite{ZXD2} \cite{ZXD3}. However, the action principle for
above connection dynamics of either
$f(R)$ theories or STT is still lacking, although the first-order action for the connection dynamics in
Einstein frame of STT was proposed in \cite{FG}.
 The purpose of this paper is to fill out this gap. We will propose
a first-order action for general STT of gravity, which includes $f(R)$ theories as special cases. The connection
dynamical formalism will be derived from this action by Hamiltonian analysis. It turns out that this connection
dynamics is exactly the same as that derived from the geometrical dynamics by canonical transformations.
Moreover, the equivalence between
this connection formalism in Jordan frame and the alternative one in Einstein frame will be proved.  Hence,
loop quantum STT, as well as loop quantum $f(R)$ theories, have got their foundation of action principle.

Throughout the  paper, we use the Latin alphabet $a$, $b$, $c$,\ldots, to represent abstract index notation
of spacetime\cite{Wald}, capital Latin alphabet $I$, $J$, $K$,\ldots,
for internal Lorentzian indices, and $i$, $j$, $k$,\ldots,
for internal $SU(2)$ indics. The other convention are as follows. The internal Minkowski metric is denoted by
$\eta_{IJ}=diag(-1,1,1,1)$. The Hodge dual of a differential form $F_{IJ}$ is denoted by ${}^\star\!F_{IJ}=\frac{1}{2}\epsilon_{IJKL}F^{KL}$, where $\epsilon_{IJKL}$
is the internal Livi-Civital symbol. The antisymmetry of a tensor $A_{IJ}$ is defined by $A_{[IJ]}=A_{IJ}-A_{JI}$.

\section{Equations of motion}\label{section1}

In order to get the Lagrangian formalism of connection dynamics of STT proposed in \cite{ZXD3},
let us first consider the following first-order action on a 4-dimensional spacetime $M$,
\begin{alignat}{1}
S[e,\omega,\phi]=&\int_M\mathcal{L}d^4x \nonumber\\
=&\int_M
\frac{1}{2}\Big(\phi ee^a_Ie^b_J\bar{\Omega}_{ab}^{~~IJ}
-2 ee^a_Ie^b_J\bar{\omega}_{a}^{IJ}\bar{\partial}_b\phi
\nonumber \\
&+ee^{[a}_Ie^{b]}_J\bar{\partial}_{a}\big( e_{b}^{I}e^{c J}\bar{\partial}_c\phi\big)
+\big(\frac{3}{2\phi}-K(\phi)\big)e\bar{\partial}_a\phi\bar{\partial}^a\phi
\nonumber \\
&-2eV(\phi)+ee^a_Ie^b_J\frac{1}{\gamma}{}^{\star}\bar{\Omega}_{ab}^{~~IJ}\Big)d^4x\,,\label{action1}
\end{alignat}
where $e=det(e^I_a)$ is the determinant of the right-handed cotetrad $e^I_a$,
$\bar{\Omega}_{ab}\!{}^{IJ}=\bar{\partial}_{[a}\bar{\omega}_{b]}^{IJ}
+\bar{\omega}_{[a}^{IK}\bar{\omega}_{b]K}^{~~~~J}$ is the curvature of the $SL(2,\mathbb{C})$ spin connection $\bar{\omega}_{a}^{IJ}$, $V(\phi)$ is the potential of the scalar field
$\phi$ with $\phi$ satisfying $\phi>0$, $K(\phi)$ is an arbitrary function of $\phi$,
and $\gamma$ is an arbitrary real number.
The variation of action (\ref{action1}) with respect to $\bar{\omega}_{a}^{IJ}$ gives
\begin{alignat}{1}
\phi\bar{\mathcal{D}}_a( e e_I^{[a}e^{b]}_J)+\frac{1}{\gamma}{}^\star\bar{\mathcal{D}}_a (e e_I^{[a}e^{b]}_J)=0\,.\label{relation}
\end{alignat}
Here the generalized derivative operator $\bar{\mathcal{D}}_a$ is defined as
\begin{alignat}{1}
\bar{\mathcal{D}}_a e^I_b=\bar{\partial}_\alpha e^I_b-\bar{\Gamma}^c_{ab}e^I_c+\bar{\omega}_a^{IJ}e_{b J}\,,
\end{alignat}
where $\bar{\Gamma}^{~~c}_{ab}$ is a torsion-free affine connection.
From Eq.(\ref{relation}) we have (see \cite{Peldan} for details)
\begin{alignat}{1}
\bar{\mathcal{D}}_{[a}( e^I_{b]})=0\,,\label{v1}
\end{alignat}
which tells us that
the spin connection $\bar{\omega}_{a}^{IJ}$ is compatible with tetrad $e^I_a$.
On the other hand, taking account of Eq.(\ref{v1}),
the variation of action (\ref{action1}) with respect to the tetrad $e^I_a$ gives
\begin{alignat}{1}
\phi G_{ab}&=
(K-\frac{3}{2\phi})((\bar{\partial}_a\phi)\,\bar{\partial}_b\phi
-\frac{1}{2}g_{ab}(\bar{\partial}_c\phi)\,\bar{\partial}^c\phi)
\nonumber \\
&+\bar{\nabla}_a\bar{\nabla}_b \phi-g_{ab}\bar{\nabla}_c\bar{\nabla}^c \phi-g_{ab}V\,,\label{v2}
\end{alignat}
where $G_{ab}$ is the Einstein tensor of $e^I_a$ and $\bar{\nabla}_a$ is
the covariant derivative operator compatible with $g_{ab}$.

Finally, taking account of Eq.(\ref{v1}),
the variation of action (\ref{action1}) with respect to the scalar field $\phi$ gives
\begin{alignat}{1}
R+2(K-\frac{3}{2\phi})\bar{\nabla}_a\bar{\nabla}^a\phi
-(K-\frac{3}{2\phi})^\prime(\bar{\partial}_a\phi)\,\bar{\partial}^a\phi-2V^\prime=0\,,\label{v3}
\end{alignat}
where a prime over a function represents a derivative with respect to the argument $\phi$.  We define a new function
\begin{alignat}{1}
\frac{\omega(\phi)}{\phi}:=K(\phi)-\frac{3}{2\phi}\,.\label{tr0}
\end{alignat}
Then it is straightforward to transform Eqs. (\ref{v2}) and (\ref{v3}) into the form in \cite{ZXD3}.
Hence the first-order action (\ref{action1}) gives exactly the equations of motion of STT.

\section{Hamiltonian analysis}\label{section2}

Let the spacetime $M$ be topologically $\Sigma\times\mathbb{R}$ for some 3-manifold $\Sigma$. One
introduces a foliation of $M$ and a time-evolution vector field $t^a$ in it.
$t^a$ can be decomposed with respect
to the unit normal vector $n^a$ of $\Sigma$ as
\begin{alignat}{1}
t^a=Nn^a+N^a\,,
\end{alignat}
where $N$ and $N^a$ are lapse function and  shift vector respectively.
In the (3+1)-decomposition of $M$, it is convenient to make a gauge fixing $n_I:=n^ae_{aI}=(1,0,0,0)$ in the internal space \cite{HM}.
In a coordinate system adopted to
the (3+1)-decomposition, the  Lagrangian density in Eq.(\ref{action1})  reads
\begin{alignat}{1}
\mathcal{L}=&\frac{1}{\gamma}\tilde{E}^b_j(\gamma \dot{K}^j_b+\dot{\omega}^j_b)
-\frac{1}{\phi}\tilde{E}^b_jK^j_b\dot{\phi}
\nonumber \\
&+\bar{K}^j_t (\mathcal{D}_b\tilde{E}^b_j-\frac{1}{\gamma\phi^2}\epsilon_{jl}^{~~m}K^l_b\tilde{E}^b_m)
\nonumber \\
&+\frac{1}{\gamma}\bar{\omega}_t^j(\partial_b\tilde{E}^b_j+\epsilon_{jl}^{~~m}(\gamma K_b^l+\omega^l_b)\tilde{E}^b_m)
\nonumber \\
&-N^a(\tilde{E}^b_j\mathcal{D}_{[a}K^j_{b]}-\frac{1}{\phi}\tilde{E}^b_jK^j_b\partial_a\phi)
\nonumber \\
&-N^a(\frac{1}{\gamma}\tilde{E}^b_j\Omega_{ab}^{~~j}-\tilde{E}^b_j\frac{1}{\gamma\phi^2}\epsilon^j_{~lm}K^l_aK^m_b)
\nonumber \\
&-\frac{\phi}{2}\underline{N}\tilde{E}^a_i\tilde{E}^b_j\epsilon^{ij}_{~~k}
(\Omega_{ab}^{~~k}-\frac{1}{\phi^2}\epsilon^k_{~lm}K^l_aK^m_b)
\nonumber \\
&-NEE^b_j(\partial_b(E^{cj}\partial_c\phi)+\omega_b^{jk}E^c_k\partial_c\phi)
\nonumber \\
&+\frac{K}{2\underline{N}}(\dot{\phi}-N^a\partial_a\phi)^2
-\frac{1}{2}(K-\frac{3}{2\phi})\underline{N}\tilde{E}^a_i\tilde{E}^{bi}(\partial_a\phi)\partial_b\phi
\nonumber \\
&+\frac{1}{\gamma}\underline{N}\tilde{E}^a_i\tilde{E}^b_j\epsilon^{ij}_{~~k}\mathcal{D}_a\omega^{k0}_b
-NEV(\phi)\,,\label{L1}
\end{alignat}
where a dot over a letter represents a derivative with respect to the time
coordinate, and we have defined
\begin{alignat}{1}
\bar{K}^i_a:=&\phi\bar{\omega}_a^{io}+\frac{1}{2}E^i_an^c\bar{\partial}_c\phi\,,\label{def1}
 \\
 \Omega_{ab}^{~~k}:=&\partial_{[a}\omega_{b]}^k+\epsilon^k_{~lm}\omega^l_a\omega^m_b\,,
\\
\bar{\omega}_a^i:=&-\frac{1}{2}\epsilon^i_{~jk}\bar{\omega}_a^{jk}\,,
\end{alignat}
and $\bar{K}^i_t:=t^a\bar{K}^i_a$, $\bar{\omega}^i_t:=t^a\bar{\omega}^i_a$ are the time component of
$\bar{K}^i_a$ and $\bar{\omega}^i_a$, E is the square root of the determinant
of the spatial metric $q_{ab}:=g_{ab}+n_an_b$, $E^a_I:=q^a_be^b_I$, $\omega_a^{IJ}:=q_a^b\bar{\omega}_b^{IJ}$
$K_a^i:=q_a^b\bar{K}_b^i$ are the spatial component of
$e^a_I$, $\bar{\omega}_a^{IJ}$ and $\bar{K}_a^i$ respectively,
 $\mathcal{D}_a$ is the spatial $SO(1, 3)$ generalized covariant
derivative operator reduced from $\bar{\mathcal{D}}_a$ and corresponds to a $SO(1, 3)$-valued
spatial connection 1-form $\omega_a^{ij}$,
$\partial_a$ is the flat derivative
operator on $\Sigma$ reduced from $\bar{\partial}_a$,
$\underline{N}:=N/E$ is the densitized lapse scalar of weight -1, and $\tilde{E}^a_i:=EE^a_i$ is
the densitized spatial triad of weight 1.

Recall that the unique torsion-free $SO(3)$ generalized
covariant derivative operator annihilating $E^a_i$ is defined as:
\begin{alignat}{1}
\nabla_aE^b_i=\partial_aE^b_i+\Gamma_{ac}^bE^b_i+\Gamma_{ai}^{~~j}E^b_j=0\,,
\label{cod}
\end{alignat}
where $\Gamma_{ac}^b$ and $\Gamma_{ai}^{~~j}$ are respectively the Levi-Civita connection and the spin
connection on $\Sigma$. For convenience we define
\begin{alignat}{1}
\Gamma_a^i:=-\frac{1}{2}\epsilon^i_{~jk}\Gamma_a^{jk}\,.
\end{alignat}
Let $C_a^i:=\omega_a^i-\Gamma_a^i$.
We further define new variables:
\begin{alignat}{1}
\gamma M_b^j:=\gamma K^j_b+C^j_b\,,\label{def2}
\\
Q_b^j:=\gamma M^j_b+\Gamma^j_b\,.
\end{alignat}
Then by using the definitions (\ref{def1}) and (\ref{def2}), the connection components $\omega^{io}_a$ can be rewritten as:
\begin{alignat}{1}
\omega_a^{io}=\frac{1}{\phi}(M^i_a-\frac{1}{\gamma}C^i_a-\frac{1}{2}E^i_an^c\bar{\partial}_c\phi)\,.
\end{alignat}
Note that we have the identity
\begin{alignat}{1}
E^b_j R_{ab}^{~~j}=0\,,
\end{alignat}
where the curvature $R_{ab}^{~~j}$ is defined as
\begin{alignat}{1}
R_{ab}^{~~j}:=\partial_{[a}\Gamma_{b]}^j+\epsilon^j_{~lm}\Gamma^l_a\Gamma^m_b\,.
\end{alignat}
Note also that the two constraint equations with respect to the Lagrangian multipliers $\bar{K}^j_t$ and $\bar{\omega}_t^j$ are equivalent to
\begin{alignat}{1}
\epsilon_{jl}^{~~m}C_b^l\tilde{E}^b_m=0\,,\label{abc}
\\
\epsilon_{jl}^{~~m}M^l_b\tilde{E}^b_m=0\,.
\end{alignat}
We will denote $\Omega^j$, $\Lambda^j$ as the corresponding Lagrangian multipliers.
Then the Lagrangian density (\ref{L1}) can be expressed as:
\begin{alignat}{1}
\mathcal{L}=&\frac{1}{\gamma}\tilde{E}^b_j\dot{Q_b^j}
-\frac{1}{\phi}\tilde{E}^b_jM^j_b\dot{\phi}
\nonumber \\
&+\Lambda^j(\partial_b\tilde{E}^b_j+\epsilon_{jl}^{~~m}Q_b^l\tilde{E}^b_m)
\nonumber \\
&-N^a(\tilde{E}^b_j\nabla_{[a}M^j_{b]}-\frac{1}{\phi}\tilde{E}^b_jM^j_b\partial_a\phi)
\nonumber \\
&-\frac{\phi}{2}\underline{N}\tilde{E}^a_i\tilde{E}^b_j\epsilon^{ij}_{~~k}
(R_{ab}^{~~k}-\frac{1}{\phi^2}\epsilon^k_{~lm}M^l_aM^m_b)
\nonumber \\
&-\underline{N}\tilde{E}^a_i\tilde{E}^{bi}\nabla_a\nabla_b\phi
\nonumber \\
&+\frac{K}{2\underline{N}}(\dot{\phi}-N^a\partial_a\phi)^2
-\frac{1}{2}(K-\frac{3}{2\phi})\underline{N}\tilde{E}^a_i\tilde{E}^{bi}(\partial_a\phi)\partial_b\phi
\nonumber \\
&-\frac{\phi}{2}\underline{N}(1+\frac{1}{\phi^2\gamma^2})(C^2-C_{ij}C^{ij})-NEV(\phi)\,,\label{L4}
\end{alignat}
where $C_{ij}:=C_{ai}\tilde{E}^a_j$ and $C:=\delta^{ij}C_{ij}$.
Since the variation of the action with respect to $C_{ij}$ gives
\begin{alignat}{1}
C_{ij}=0\,,
\end{alignat}
the Lagrangian density (\ref{L4}) can be reduced to
\begin{alignat}{1}
\mathcal{L}=&\frac{1}{\gamma}\tilde{E}^b_j\dot{A_b^j}
-\frac{1}{\phi}\tilde{E}^b_jK^j_b\dot{\phi}
\nonumber \\
&+\Lambda^j(\partial_b\tilde{E}^b_j+\epsilon_{jl}^{~~m}A_b^l\tilde{E}^b_m)
\nonumber \\
&-N^a(\tilde{E}^b_j\nabla_{[a}K^j_{b]}-\frac{1}{\phi}\tilde{E}^b_jK^j_b\partial_a\phi)
\nonumber \\
&-\frac{\phi}{2}\underline{N}\tilde{E}^a_i\tilde{E}^b_j\epsilon^{ij}_{~~k}
(R_{ab}^{~~k}-\frac{1}{\phi^2}\epsilon^k_{~lm}K^l_aK^m_b)
\nonumber \\
&+\frac{K}{2\underline{N}}(\dot{\phi}-N^a\partial_a\phi)^2
-\frac{1}{2}(K-\frac{3}{2\phi})\underline{N}\tilde{E}^a_i\tilde{E}^{bi}(\partial_a\phi)\partial_b\phi
\nonumber \\
&-\underline{N}\tilde{E}^a_i\tilde{E}^{bi}\nabla_a\nabla_b\phi-NEV(\phi)\,,
\label{L5}
\end{alignat}
where
\begin{alignat}{1}
A_b^j:=\gamma K^j_b+\Gamma^j_b\,.
\end{alignat}
By Legendre transformation, the momentum conjugate to the configuration variables $A^i_a$
and $\phi$ are defined respectively as
\begin{alignat}{1}
&\pi^a_i:=\frac{\delta\mathcal{L}}{\delta\dot{A_a^i}}=\frac{1}{\gamma}\tilde{E}^a_i\,,
\\
&\pi:=\frac{\delta\mathcal{L}}{\delta\dot{\phi}}=-\frac{1}{\phi}\tilde{E}^b_jK^j_b+
\frac{K}{\underline{N}}(\dot{\phi}-N^a\partial_a\phi)\,.\label{pi}
\end{alignat}
The fundamental Poisson brackets read
\begin{alignat}{1}
\big\{A_a^i(x)\,,\tilde{E}^b_j(y)\big\}
&=\gamma\delta^b_a\delta^i_j\delta^3(x-y)\,,
\\
\big\{\phi(x)\,,\pi(y)\big\}&=\delta^3(x-y)\,.
\end{alignat}
It should be noted that the second-class constraints appeared in the Hamiltonian analysis have been solved by the partial gauge fixing.
In the case when $K\neq0$, the corresponding Hamiltonian reads
\begin{alignat}{1}
H=\int d^3x(\Lambda^i\mathcal{G}_i
+N^a\mathcal{C}_a+\underline{N}\mathcal{C})\,,
\end{alignat}
where the  Gaussian,
vector and scalar constraints read respectively as:
\begin{alignat}{1}
\mathcal{G}_j=&\partial_b\tilde{E}^b_j+\epsilon_{jl}^{~~m}A_b^l\tilde{E}^b_m\,,
\\
\mathcal{C}_a=&\tilde{E}^b_j\nabla_{[a}K^j_{b]}+\pi\partial_a\phi\,,
\\
\mathcal{C}=&\frac{\phi}{2}\tilde{E}^a_i\tilde{E}^b_j\epsilon^{ij}_{~~k}
(R_{ab}^{~~k}-\frac{1}{\phi^2}\epsilon^k_{~lm}K^l_aK^m_b)
\nonumber \\
&+\tilde{E}^a_i\tilde{E}^{bi}\nabla_a\nabla_b\phi
+\frac{1}{2}(K-\frac{3}{2\phi})\tilde{E}^a_i\tilde{E}^{bi}(\partial_a\phi)\partial_b\phi
\nonumber \\
&+\frac{1}{2K}(\pi+\frac{1}{\phi}\tilde{E}^b_jK^j_b)^2+E^2V(\phi)\,.
\end{alignat}
In the special case when $K=0$, it is easy to see from Eq.(\ref{pi}) that there is a primary constraint
\begin{equation}
S=\pi\phi+\tilde{E}^b_jK^j_b\,,
\end{equation}
which is called
the conformal constraint in \cite{ZXD3}. Thus the Hamiltonian becomes
\begin{alignat}{1}
H=\int d^3x(\Lambda^i\mathcal{G}_i
+N^a\mathcal{C}_a+\underline{N}\mathcal{C}_0+\lambda S)\,,
\end{alignat}
where the scalar constraint reads
\begin{alignat}{1}
\mathcal{C}_0=&\frac{\phi}{2}\tilde{E}^a_i\tilde{E}^b_j\epsilon^{ij}_{~~k}
(R_{ab}^{~~k}-\frac{1}{\phi^2}\epsilon^k_{~lm}K^l_aK^m_b)
\nonumber \\
&+\tilde{E}^a_i\tilde{E}^{bi}\nabla_a\nabla_b\phi
-\frac{3}{4\phi}\tilde{E}^a_i\tilde{E}^{bi}(\partial_a\phi)\partial_b\phi
\nonumber \\
&+E^2V(\phi).
\end{alignat}
It is obvious that the above Hamiltonian formulations in both cases coincide with those in \cite{ZXD3}.

On the other hand, as pointed out in \cite{FG}, the following first-order action
\begin{alignat}{1}
S[e,\omega,\phi]&=\int\Big[\frac{1}{2}\phi e e_I^{a} e_J^{b} \big(\bar{\Omega}_{ab} \!{}^{IJ}+\frac{1}{\gamma}{}^\star\!\bar{\Omega}_{ab }\!{}^{IJ}\big)
\nonumber \\
&-\frac{1}{2}K(\phi)ee^{Ia}e^b_I(\bar{\partial}_a\phi)\,\bar{\partial}_b\phi
-e V(\phi)\Big]d^4x\,,
\label{action10}
\end{alignat}
can give a connection dynamics of STT in Einstein frame. We now show that the Hamiltonian formalism of action (\ref{action10})
is equivalent to the one which we just derived from action (\ref{action1}), because they are related
to each other by
a canonical transformation. In the case when $K\neq0$, the
  Hamiltonian corresponding to action (\ref{action10}) is
a linear combination of first-class constraints as
\begin{alignat}{1}
H=\int d^3x(\Lambda^i\hat{\mathcal{G}}_i
+N^a\hat{\mathcal{C}}_a+\underline{N}\hat{\mathcal{C}})\,,
\end{alignat}
where
\begin{alignat}{1}
\hat{\mathcal{G}}_i&=\gamma^{-1}\hat{\mathcal{D}}_a\hat{E}^a_i\,,\label{eq:c1}
\\
\hat{\mathcal{C}}_a&=\hat{E}^b_i\hat{F}_{ab}^{~~i}+\hat{\pi}\partial_a\phi\,,\label{eq:c2}
\\
\hat{\mathcal{C}}&=-\gamma^{-1}\frac{1}{2\phi}\epsilon^{ij}_{~~k}\hat{E}^a_i\hat{E}^b_j[\hat{F}_{ab}^{~~k}
-(\gamma+\gamma^{-1})\hat{R}_{ab}^{~~k}]
\nonumber \\
&+\frac{K(\phi)}{2\phi^2}\hat{E}^{ai}\hat{E}^{b}_i(\partial_a\phi)\,\partial_b\phi+\frac{\hat{\pi}^2}{2K(\phi)}
\nonumber \\
&+V\sqrt{\det(\tilde{E}^{ai}\tilde{E}^b_i)}\,,
\label{eq:c3}
\end{alignat}
with
\begin{alignat}{1}
\hat{\mathcal{D}}_a\hat{E}^a_i:=\partial_a\hat{E}^a_i+\gamma\epsilon_{ij}^{~~k}\hat{A}_a^j\hat{E}^a_k\,,
\end{alignat}
and $\hat{F}_{ab}^{~~i}$ and $\hat{R}_{ab}^{~~i}$ standing for the curvature of $\hat{A}_a^i$ and $\hat{\Gamma}_a^i$ respectively, i.e.,
\begin{alignat}{1}
\hat{F}_{ab}^{~~i}&=\partial_{[a}\hat{A}_{b]}^i+\gamma\epsilon^i_{~jk}\hat{A}_a^j\hat{A}_b^k\,,
\label{cu1}
\\
\hat{R}_{ab}^{~~i}&=\partial_{[a}\hat{\Gamma}_{b]}^i+\epsilon^i_{~jk}\hat{\Gamma}_a^j\hat{\Gamma}_b^k\,.
\label{cu2}
\end{alignat}
Here $\hat{\Gamma}_a^i$ is the $SU(2)$ spin connection satisfying
\begin{alignat}{1}
\hat{D}_a\hat{E}^b_i=\partial_a\hat{E}^b_i+\hat{\Gamma}_{ac}^{\;\;\;b}\hat{E}^c_i
-\hat{\Gamma}_{ca}^{\;\;\;c}\hat{E}^b_i
+\epsilon_{ij}^{~~k}\hat{\Gamma}_a^j\hat{E}^b_k=0\,,
\end{alignat}
where $\hat{\Gamma}_{ab}^{\;\;\;c}$ is the Christoffel connection determined by the
spatial metric
\begin{alignat}{1}
\hat{q}^{ab}=\hat{E}\hat{E}^{ai}\hat{E}^{bi}\,,
\end{alignat}
with $\hat{E}:=1/det(\hat{E}^a_i)$.
The fundamental Poisson brackets are
\begin{alignat}{1}
\big\{\hat{A}_a^i(x)\,,\hat{E}^b_j(y)\big\}
&=\delta^b_a\delta^i_j\delta^3(x-y)\,,
\label{eq:p3}
\\
\big\{\phi(x)\,,\hat{\pi}(y)\big\}&=\delta^3(x-y)\,.
\label{eq:p4}
\end{alignat}
To do  the canonical transformation, we
first define
\begin{alignat}{1}
K_a^i:&=\phi(\hat{A}_a^i-\gamma^{-1}\hat{\Gamma}_a^i)\,,
\label{eq:tr1}
\\
\tilde{E}^a_i:&=\phi^{-1}\hat{E}^a_i\,.
\end{alignat}
Then we further define
\begin{alignat}{1}
\pi:&=\hat{\pi}-\frac{1}{\phi}K_a^i\tilde{E}^a_i\,,
\\
A_a^i:&=\Gamma_a^i+\gamma K_a^i\,.
\end{alignat}
Using Eqs. (\ref{eq:p3}) and (\ref{eq:p4}), we can get the Poisson brackets between new variables as
\begin{alignat}{1}
\big\{A_a^i(x)\,,\tilde{E}^b_j(y)\big\}
&=\gamma\delta^b_a\delta^i_j\delta^3(x-y)\,,
\label{eq:p5}
\\
\big\{\phi(x)\,,\pi(y)\big\}&=\delta^3(x-y)\,,
\label{eq:p6}
\\
\big\{A_a^i(x)\,,A_b^j(y)\big\}
&=0=\big\{\tilde{E}^a_i(x)\,,\tilde{E}^b_j(y)\big\}\,,
\\
\big\{\phi(x)\,,\phi(y)\big\}
&=0=\big\{\pi(x)\,,\pi(y)\big\}\,.
\end{alignat}
Taking account of Eq.(\ref{tr0}), the constraints (\ref{eq:c1}), (\ref{eq:c2}) and (\ref{eq:c3})
can be written in terms of new variables, up to Gaussian constraint, as
\begin{alignat}{1}
\hat{\mathcal{G}}_i=&\gamma(\partial_a\tilde{E}^a_i+\epsilon_{ij}^{~~k}A_a^j\tilde{E}^a_k)\,,
\label{c11}
\\
\hat{\mathcal{C}}_a=&\gamma^{-1}\tilde{E}^b_i F_{ab}^{~~i}+\pi\partial_a\phi\,,
\label{c22}
\\
\hat{\mathcal{C}}=&\frac{\phi}{2}\epsilon_i^{~lm}\tilde{E}^a_l\tilde{E}^b_m[F_{ab}^{~~i}
-(\gamma^{2}+\frac{1}{\phi^2})\epsilon^i_{~jk}K_a^jK_b^k]
\nonumber \\
&+\frac{1}{2\,\omega(\phi)+3}\bigg(\frac{1}{\phi}(K_a^i\tilde{E}^a_i)^2
+2\tilde{\pi}K_a^i\tilde{E}^a_i+\pi^2\phi\bigg)
\nonumber \\
&+\frac{\omega(\phi)}{2\phi}\tilde{E}^{ai}\tilde{E}^b_i
(\partial_a\phi)\,\partial_b\phi
+\tilde{E}^{ai}\tilde{E}^{b}_i(\partial_a\,\partial_b\phi-
\Gamma_{ab}^{\;\;\;c}\partial_c\phi)
\nonumber \\
&+V\sqrt{\det(\tilde{E}^{ai}\tilde{E}^b_i)}\,,
\label{c33}
\end{alignat}
where $F_{ab}^{~~i}:=\partial_{[a}A_{b]}^i+\epsilon^i_{~jk}A_a^jA_b^k$.
It is obvious that these  constraints coincide with our results as well as those in \cite{ZXD3}.
Similarly, it is easy to get the same conclusion in the special case when $K=0$.

\section{concluding remarks}\label{section5}

As candidate modified gravity theories, STT provide the great possibility to account for
the dark Universe and some fundamental issues in physics. The nonperturbative loop quantization of
STT is based on their connection dynamical formalism obtained in Hamiltonian formulation in \cite{ZXD3}.
The achievement in this paper is to set up an action principle for the connection dynamics of STT in Jordan frame.
Since $f(R)$ theories of gravity can be regarded as the  special kinds of STT, our action principle is
also valid for the connection dynamics of $f(R)$ theories. To get the action principle, we first show
that the first-order action (\ref{action1}) gives the right equations of motion for general STT.
Then a detailed Hamiltonian analysis is done to this action. By a partial gauge fixing, the internal
$SL(2,\mathbb{C})$ group of the theory is reduced to $SU(2)$, and the second-class constraints are solved.
Thus we obtain a first-class Hamiltonian system with a $SU(2)$ connection as a configuration variable.
This Hamiltonian formalism is exactly the same as the one in \cite{ZXD3} derived from the
geometrical dynamics by canonical transformations.

On the other hand, the directly corresponding Hamiltonian connection formulation of action (\ref{action10}) is in Einstein
frame, while as shown in \cite{ZXD3} the natural connection formulation obtained by canonical transformations
in Hamiltonian framework is in Jordan frame. However we have shown that they are equivalent to each other at classical
level. Nevertheless, the ambiguity, whether one should start with the Jordan frame or Einstein frame to
quantize STT, still exits.
Besides providing the action principle for connection dynamics of STT, actions (\ref{action1}) and (\ref{action10}) also
lay the foundation of spinfoam path-integral quantization of STT. We leave this issue for future study.

\begin{acknowledgements}

This work is supported by NSFC (No. 10975017 and No. 11235003) and the Fundamental
Research Funds for the Central Universities.
\end{acknowledgements}


\newcommand\AL[3]{~Astron. Lett.{\bf ~#1}, #2~ (#3)}
\newcommand\AP[3]{~Astropart. Phys.{\bf ~#1}, #2~ (#3)}
\newcommand\AJ[3]{~Astron. J.{\bf ~#1}, #2~(#3)}
\newcommand\APJ[3]{~Astrophys. J.{\bf ~#1}, #2~ (#3)}
\newcommand\APJL[3]{~Astrophys. J. Lett. {\bf ~#1}, L#2~(#3)}
\newcommand\APJS[3]{~Astrophys. J. Suppl. Ser.{\bf ~#1}, #2~(#3)}
\newcommand\JCAP[3]{~JCAP. {\bf ~#1}, #2~ (#3)}
\newcommand\LRR[3]{~Living Rev. Relativity. {\bf ~#1}, #2~ (#3)}
\newcommand\MNRAS[3]{~Mon. Not. R. Astron. Soc.{\bf ~#1}, #2~(#3)}
\newcommand\MNRASL[3]{~Mon. Not. R. Astron. Soc.{\bf ~#1}, L#2~(#3)}
\newcommand\NPB[3]{~Nucl. Phys. B{\bf ~#1}, #2~(#3)}
\newcommand\PLB[3]{~Phys. Lett. B{\bf ~#1}, #2~(#3)}
\newcommand\PRL[3]{~Phys. Rev. Lett.{\bf ~#1}, #2~(#3)}
\newcommand\PR[3]{~Phys. Rep.{\bf ~#1}, #2~(#3)}
\newcommand\PRD[3]{~Phys. Rev. D{\bf ~#1}, #2~(#3)}
\newcommand\SJNP[3]{~Sov. J. Nucl. Phys.{\bf ~#1}, #2~(#3)}
\newcommand\ZPC[3]{~Z. Phys. C{\bf ~#1}, #2~(#3)}
\newcommand\CQG[3]{~Class. Quant. Grav. {\bf ~#1}, #2~(#3)}

\end{document}